\documentclass[%
 reprint,
 amsmath,amssymb,
 aps,
]{revtex4-2}

\usepackage{graphicx}
\usepackage{dcolumn}
\usepackage{bm}
\usepackage{algorithm}
\usepackage[noend]{algpseudocode}

\usepackage{amsmath}
\usepackage{amssymb}
\usepackage{color}
\usepackage{amsthm}

\theoremstyle{definition}

\newtheorem{theorem}{Theorem}

{
	\theoremstyle{plain}
	\newtheorem{assumption}{Condition}
}

\begin{document}

\preprint{APS/123-QED}

\title{Exact Crystalline Structure Recovery in X-ray Crystallography from Coded Diffraction Patterns}

\author{Samuel Pinilla}
 \email{samuel.pinilla@correo.uis.edu.co}
 \homepage{http://diffraction.uis.edu.co}
 \affiliation{Department of Electronic Engineering, Universidad Industrial de Santander}
\author{Jorge Bacca}%
\email{jorge.bacca1@correo.uis.edu.co}
\affiliation{Department of Computer Science, Universidad Industrial de Santander}%

\author{Cesar Vargas}
\affiliation{ Fundaci\'on Universitaria Konrad Lorenz}%
\author{Juan Poveda}
\affiliation{Department of Chemistry, Universidad Industrial de Santander}

\author{Henry Arguello}
\affiliation{Department of Computer Science., Universidad Industrial de Santander}

\date{\today}

\begin{abstract}
X-ray crystallography (XC) is an experimental technique used to determine three-dimensional crystalline structures. The acquired data in XC, called diffraction patterns, is the Fourier magnitudes of the unknown crystalline structure. To estimate the crystalline structure from its diffraction patterns, we propose to modify the traditional system by including an optical element called coded aperture which modulates the diffracted field to acquire coded diffraction patterns (CDP). For the proposed coded system, in contrast with the traditional, we derive exact reconstruction guarantees for the crystalline structure from CDP (up to a global shift phase). Additionally, exploiting the fact that the crystalline structure can be sparsely represented in the Fourier domain, we develop an algorithm to estimate the crystal structure from CDP. We show that this method requires 50\% fewer measurements to estimate the crystal structure in comparison with its competitive alternatives. Specifically, the proposed method is able to reduce the exposition time of the crystal, implying that under the proposed setup, its structural integrity is less affected in comparison with the traditional. We discuss further implementation of imaging devices that exploits this theoretical coded system.
\end{abstract}

\maketitle

\section{Introduction}
\begin{figure}[t]
	\centering\includegraphics[width = 0.75\linewidth]{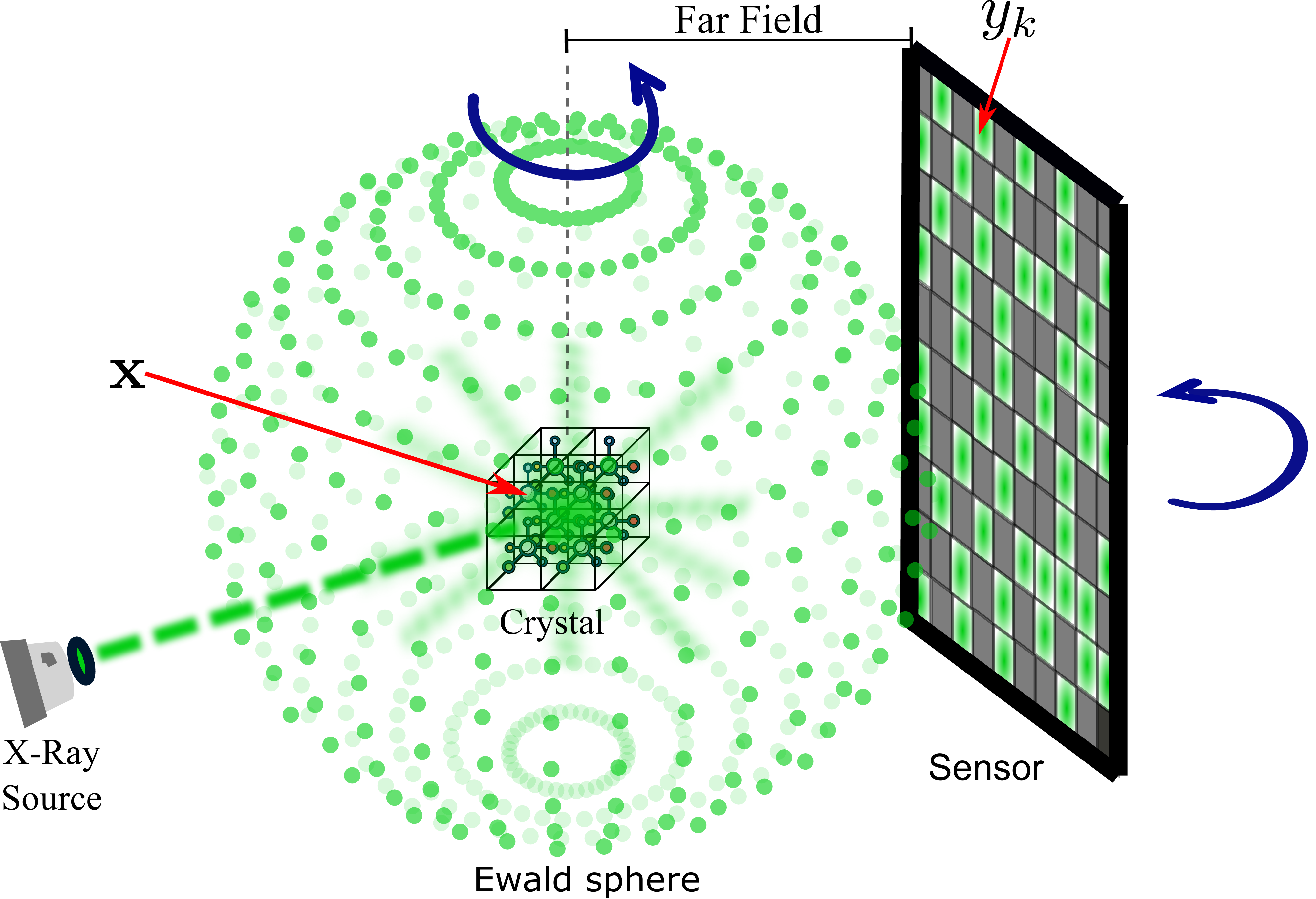}
	\caption{Traditional acquisition system of diffraction patterns in XC. An X-ray source irradiates a crystal producing diffraction patterns that are recorded with a two-dimensional sensor.}
	\label{fig:traditional}
	\vspace{-1.5em}
\end{figure}

X-ray Crystallography (XC) is known as the leading technique for molecular structure characterization in material analysis \cite{xcrys}. Specifically, XC plays an essential role in fields as biology \cite{biology1978apply}, drug design \cite{drugdesign2003application}, and material sciences \cite{nano2016apply}, among others. The traditional acquisition system in XC is illustrated in Fig.\ref{fig:traditional}, where the acquired data forms ``the Ewald sphere'' according to the perturbation theory which is mathematically modeled as the Fourier magnitude of the unknown crystalline structure called diffraction patterns \cite{hermann2017crystallography}. Specifically, in Fig.\ref{fig:traditional} the resulting diffraction patterns, when an X-ray source irradiates a crystal, are recorded while both the sensor and the crystal gradually rotate~\cite{smyth2000x}. 

The crystal structure can be uniquely estimated from the phase of its diffraction patterns \cite{hermann2017crystallography}. Although, the phase of the diffraction patterns cannot be directly measured, it can be recovered from their phaseless intensity measurements \cite{7078985}, problem that is known as phase retrieval (PR) \cite{7078985,kim1991phase}. Mathematically, the PR problem in XC consists on retrieving a discrete version of the crystalline structure $\mathbf{x}\in \mathbb{C}^{n}$ from the phaseless measurements $y_{k} = \lvert \mathbf{f}_{k}^{H}\mathbf{x}\rvert^{2}$ where $n$ is the discrete size of the crystal, $k=1,\cdots,m$ with $m$ the total number of measurements, $\mathbf{f}_{k}\in \mathbb{C}^{n}$ are the vectors that models the three-dimensional discrete Fourier transform \cite{shechtman2015phase}, and $(\cdot)^{H}$ represents the conjugate transpose operation. The PR problem has been challenging due to its infinite solutions \cite{7078985,bendory2017fourier}. Specifically, there are so-called trivial ambiguities that are always present \cite{7078985}. The following three transformations (or any combination of them) conserve Fourier magnitude: 1) \textit{global shift phase:} $\mathbf{x}[a] = \mathbf{x}[a] e^{j\phi_{0}}$; 2) \textit{conjugate inversion}: $\mathbf{x}[a] = \overline{\mathbf{x}[-a]}$; 3) \textit{spatial shift}: $\mathbf{x}[a] = \mathbf{x}[a + a_{0}]$. Moreover, the number of measurements $m$ must satisfy $m\geq 4n-1$ to accurately retrieve the signal $\mathbf{x}$ (up to trivial ambiguities) \cite{conca2015algebraic,bendory2017fourier}. i.e. a 4-fold time exposition X-ray radiation is required, which leads to degradation of the crystalline structure \cite{chapman2011femtosecond,lomb2011radiation,yano2005x,owen2006experimental,henderson1995potential,riekel2004recent}. Additionally, if the crystal deteriorates under this irradiation, material characterization becomes more challenging \cite{yano2005x}.

Despite such challenges, several methods have been develop to retrieve the crystalline structure. To name a few, the error reduction method \cite{fienup1982phase}, dual space programs \textit{SnB} and \textit{ShelxD} \cite{sheldrick2008short}, iterated projections \cite{elser2003solution}, and the charge flipping algorithm \cite{oszlanyi2004ab}. All these methods alternate between real and reciprocal space by the Fourier transform by imposing constraints on the real-space charge density. All these algorithms require to over-expose the crystal to the X-ray source in order to acquired the large enough amount of diffraction patterns needed to estimate the crystalline structure i.e. they work under the $m\geq 4n-1$ regime. Further, these approaches tend to return inaccurate estimates of the crystalline structure when the measurements $y_{k}$ are corrupted by noise \cite{7078985}. Also, the convergence to the true crystalline structure of the above methods is not guaranteed.

This work proposes to modify the traditional system by including an optical element called coded aperture which modulates the diffracted field to acquire coded diffraction patterns (CDP). The main advantage of the proposed system is that only the global phase shift ambiguity appears, that is, the conjugate inversion and spatial shift ambiguities will not occur \cite{cande1,pinilla2018coded1}. This implies that retrieving the crystalline structure from CDP is much effortless compared with the traditional XC. For the proposed coded system, we derive exact reconstruction guarantees for the crystalline structure from CDP (up to a global shift phase). Additionally, given the fact that the crystalline structure is a periodic element, it can be sparsely represented in the Fourier domain, i.e. the number of non-zero coefficients of the Fourier transformed crystal is much smaller than $n$. Thus, exploiting the sparsity prior of the crystal we develop an algorithm to estimate the crystalline structure from CDP  that requires 50\% fewer measurements in comparison with its competitive alternatives. Specifically, the proposed method is able to reduce the exposition time of the crystal, implying that under the proposed setup, its structural integrity is less affected in comparison with the traditional. Numerical simulations are conducted to evaluate the performance of the proposed method using synthetic data.


\section{Problem Formulation}
\label{sec:problem}

\begin{figure}[h]
	\centering\includegraphics[width = 0.75\linewidth]{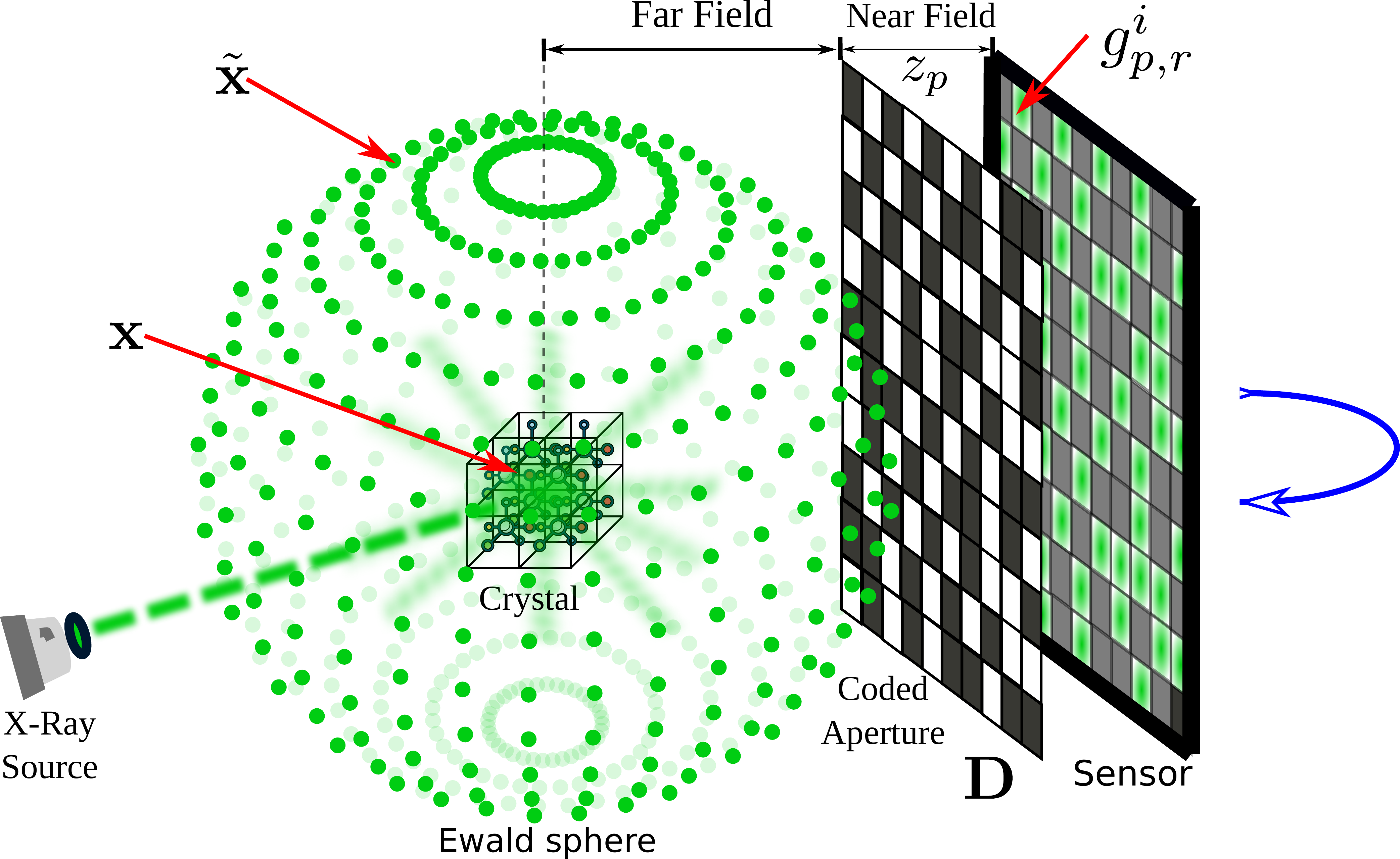}
	\caption{Illustrative configuration to acquire coded diffraction patterns from a crystal. A coded aperture is located at a distance $z_{p}$ from the sensor such that the acquired data corresponds with diffraction patterns in the near field.}
	\label{fig:proposed}
\end{figure}

In contrast to the traditional acquisition system illustrated in Fig. \ref{fig:traditional}, a coded diffraction system, as shown in Fig. \ref{fig:proposed}, is proposed. Notice that Fig. \ref{fig:proposed} includes an optical element known as coded aperture, modeled as $\mathbf{D}$, at distance $z_{p}$ from the sensor. This optical element modules the diffracted field before being measured by the sensor. Besides, changing the distance $z_{p}$, this acquisition system allows acquiring multiple snapshots, where $p=1,\cdots,P$ indexes the sensing distances. The signal $\tilde{\mathbf{x}}\in \mathbb{C}^{n}$ in Fig. \ref{fig:proposed} corresponds with the Fourier transform of $\mathbf{x}$, that is, $\tilde{\mathbf{x}} = \mathbf{F}\mathbf{x}$, where $\mathbf{F}$ represents the 3D discrete Fourier transform matrix.


\begin{figure}[h]
	\centering\includegraphics[width=1\linewidth]{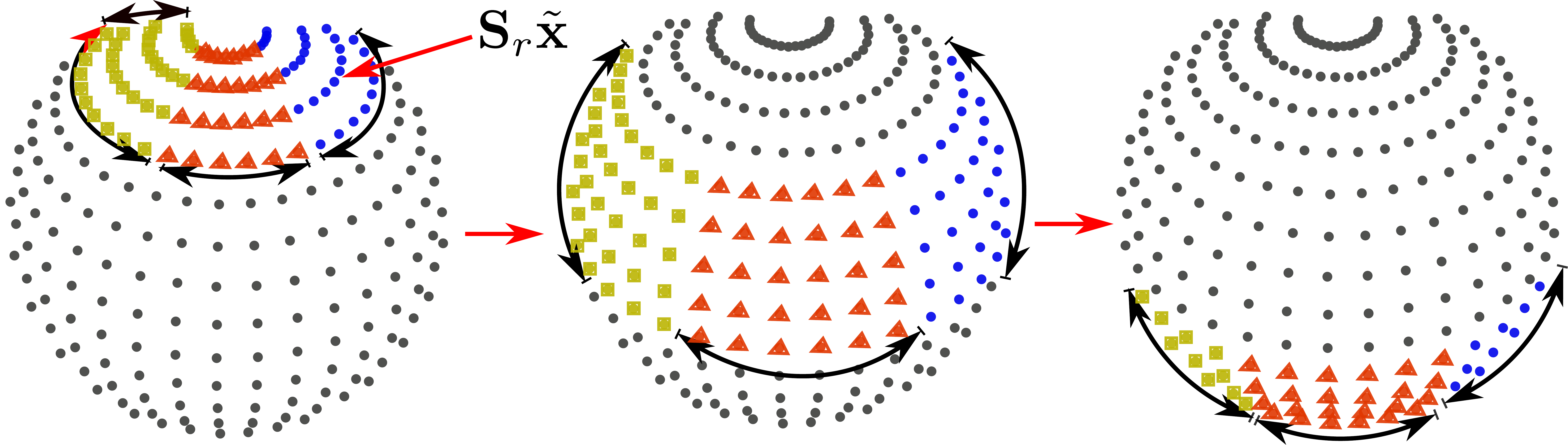}
	\caption{Illustration of the proposed scanning process to ``the Ewald sphere'' of the system in Fig. \ref{fig:proposed}. Each highlighted region in the dashed sphere corresponds to a different rotation of the sensor and the coded aperture.}
	\label{fig:scan}
\end{figure}

Observe that in Fig. \ref{fig:proposed}, we assume that the coded aperture $\mathbf{D}$ is located at the far field from the crystal. Moreover, the distance $z_{p}$ between the coded aperture and the sensor is chosen so that the coded diffracted field, captured at the detector, belongs to the near field. Additionally, we assume that the coded aperture and the sensor jointly perform a raster scanning across ``the Ewald sphere'', as illustrated in Fig. \ref{fig:scan}, while the crystal remains fixed. Thus, in contrast with the traditional architecture, crystal rotation is avoided. Mathematically, a given highlighted region in Fig. \ref{fig:scan} is modeled as $\tilde{\mathbf{x}}_{r}=\mathbf{S}_{r}\tilde{\mathbf{x}}$ where $\mathbf{S}_{r}$ is a selection diagonal matrix where $r=1,\cdots,R$ with $R$ as the number of regions. 

In order to easily model the acquired data in Fig. \ref{fig:proposed}, $\hat{\mathbf{D}}\in \mathbb{C}^{n\times n}$ is defined as a diagonal matrix whose entries are the elements of $\mathbf{D}$. Thus, from the traditional diffraction theory \cite{pinilla2018coded1,poon2014introduction}, the measurements being captured at the detector for the $r$-th region and at a distance $z_{p}$, are given by
\begin{align}
	g^{i}_{p,r} = \lvert \langle\mathbf{a}_{p,i},\tilde{\mathbf{x}}_{r}\rangle \rvert^{2}, i=1\cdots,n,
	\label{eq:codedmeas}
\end{align}
where $\mathbf{a}_{p,i}$ are the sampling vectors defined as
\begin{align}
	\mathbf{a}_{p,i} = \overline{\hat{\mathbf{D}}}\mathbf{F}\overline{\mathbf{T}}(z_{p})\mathbf{f}_{u_{i}},
	\label{eq:samplingvectors}
\end{align}
where $u_{i} = (i-1)\mod n +1$, $\mathbf{T}(z_{p})$ is the spatial frequency transfer function which depends on the distance $z_{p}$ \cite{pinilla2018coded1,poon2014introduction}, and $\overline{(\cdot)}$ represents the conjugate operation. Now, considering the definition of $\tilde{\mathbf{x}}_{r}$ we have that \eqref{eq:codedmeas} can be equivalently expressed as
\begin{align}
	g^{i}_{p,r} = \lvert \langle\mathbf{b}^{r}_{p,i},\mathbf{x}\rangle \rvert^{2}, i=1\cdots,n,
	\label{eq:modelfinal}
\end{align}
where $\mathbf{b}^{r}_{p,i} = \mathbf{F}^{H}\mathbf{S}_{r}\mathbf{a}_{p,i}$. Observe that under the proposed acquisition system, the maximum number of measurements is given by $m\leq nRP$. Additionally, we remark that the acquired measurements ${g^{i}_{p,r}}$ in Fig. \ref{fig:proposed} are different than ${y_{k}}$ in Fig. \ref{fig:traditional} due to the inclusion of the coded aperture.

This paper deals with the problem of recovering $\mathbf{x}\in \mathbb{C}^{n}$ from the phaseless measurements $\{g^{i}_{p,r}\}$. Additionally, we assume that the entries of a coded aperture $\mathbf{D}$ are $i.i.d$ copies of a discrete random variable $d$ obeying  $|d|\leq 1$. This assumption over the coded aperture is important because under a random sensing process uniqueness is guaranteed (up to a global phase shift) as will be shown in Section \ref{sec:gurantees}. This implies that retrieving the crystalline structure from CDP is much easier compared with the traditional XC since the conjugate inversion and spatial shift ambiguities will not occur \cite{cande1,pinilla2018coded1}. Moreover, exploiting the fact that the crystalline structure is periodic, it can be sparsely represented in the Fourier domain, i.e. $\lVert \mathbf{Fx} \rVert_{0}=s \ll n$ where $\lVert\cdot \rVert_{0}$ is the $\ell_{0}$ pseudo-norm, we formulate the following optimization problem
\begin{align}
\hspace{-8em}\min_{\mathbf{x}\in \mathbb{C}^{n}} \hspace{1em} h(\mathbf{x}) &= \frac{1}{m}\sum_{i=1}^{n}\sum_{r=1}^{R}\sum_{p=1}^{P}\left(\sqrt{g^{i}_{p,r}} - \lvert \langle\mathbf{b}^{r}_{p,i},\mathbf{x}\rangle \rvert\right)^{2},
\nonumber\\
\text{ s.t }\hspace{1em}\lVert\mathbf{Fx} \rVert_{0}&\leq s
\label{eq:optiproblem}
\end{align}
To solve \eqref{eq:optiproblem} we propose a gradient descend method that will be described in Section \ref{sec:method}.

\section{Exact Recovery Guarantees}
\label{sec:gurantees}
Observe that \eqref{eq:modelfinal} can be equivalently expressed as
\begin{align}
	g^{i}_{p,r} = (\mathbf{b}^{r}_{p,i})^{H}\mathbf{x}\mathbf{x}^{H}\mathbf{b}^{r}_{p,i}.
	\label{eq:problem}
\end{align}
Now, consider the linear operator $\mathcal{B}:\mathcal{S}^{n\times n}\rightarrow \mathbb{R}^{m}$($\mathcal{S}^{n\times n}$ is the space of self-adjoint matrices) defined as
\begin{equation}
	\small
	\mathcal{B}(\mathbf{W})=\left[(\mathbf{b}^{1}_{1,1})^{H}\mathbf{W}\mathbf{b}^{1}_{1,1},\cdots,(\mathbf{b}^{R}_{P,n})^{H}\mathbf{W}\mathbf{b}^{R}_{P,n}\right]^{T},
	\label{eq:map}
\end{equation}
and stacking the measurements $\{ g^{i}_{p,r}\}$ as $\mathbf{g}:=[g^{1}_{1,1},\cdots, g^{n}_{P,R}]^{T}$, then we have that
\begin{align}
	\mathbf{g} = \mathcal{B}(\mathbf{x}\mathbf{x}^{H}).
	\label{eq:operator}
\end{align}
Thus, to prove that the signal $\mathbf{x}$ can be exactly recovered from the measurements $g^{i}_{p,r}$ in \eqref{eq:problem}, from \eqref{eq:operator} we have that $\mathcal{B}(\cdot)$ must be injective \cite{candes2013phase,candes2013phaselift}. More precisely, this work follows the strategy in \cite{gross2017improved} that considers 
\begin{align}
\mathcal{T}_{\mathbf{x}} = \left\lbrace \mathbf{x}\mathbf{w}^{H} + \mathbf{w}\mathbf{x}^{H}| \mathbf{w}\in \mathbb{C}^{n} \right \rbrace,
\label{eq:tangent}
\end{align}
as the tangent space of the manifold of all rank-1 Hermitian matrices at the point $\mathbf{x}\mathbf{x}^{H}$. Thus, if the operator $\mathcal{B}$ satisfies the following condition, which is proved in Theorem~\ref{theo:guarantee}, one can guarantee recovery with high probability\cite{candes2013phase}.
\begin{assumption}
	For any $\delta\in(0,1)$ and some constant $\beta>0$ the linear operator $\mathcal{B}$ satisfies 
	\begin{equation}
		(1-\delta)\|\mathbf{W}\|_{1}\leq\frac{1}{\beta}\|\mathcal{B}(\mathbf{W})\|_{1}\leq(1+\delta)\|\mathbf{W}\|_{1},
		\label{rip}
	\end{equation} 
	for all matrices $\mathbf{W}\in \mathcal{T}_{\mathbf{x}}$, where  $\|\mathbf{W}\|_1= \sum_{i}\sigma_i(\mathbf{W})$ with $\sigma_i(\mathbf{W})$ as the $i$-th singular value of $\mathbf{W}$.
	\label{assu:assumption}
\end{assumption}

\begin{theorem}
	Fix any $\delta\in(0,1)$ and the coded aperture $\hat{\mathbf{D}}\in \mathbb{C}^{n\times n}$, with $i.i.d$ copies of a random variable $d$ such that $\lvert d\rvert\leq 1$. Then, we have that
	\begin{align}
		\mathcal{P}\left(\frac{1}{m}\lVert \mathbf{B} \rVert^{2}_{\infty} \leq 1+\delta\right)\leq 1-2e^{-c_{0}m\epsilon^{2}},
		\label{eq:probability}
	\end{align}
	for some constant $c_{0}>0$ provided that $m\geq C s$ where $s$ is the sparsity of the crystalline structure with $C>0$ and $\mathbf{B}$ is given by $\mathbf{B} = \left[\mathbf{b}^{r}_{1,1},\cdots,\mathbf{b}^{R}_{P,n}\right]^{H}$. Also, Condition \ref{assu:assumption} is satisfied with the same probability taking $\beta = m$, where $\lVert \mathbf{B} \rVert_{\infty} $ denotes the spectral norm of $\mathbf{B}$.
	\label{theo:guarantee}
\end{theorem}
The proof is deferred to Appendix \ref{app:proofTheogua}. Theorem \ref{theo:guarantee} establishes two aspects: first, the only ambiguity that appears in the proposed system is the global phase shift. Second, the condition $m\geq 4n-1$ imposed by the traditional system in XC can be defeated by the proposed architecture since now the number of measurements $m$ depends on the sparsity ($s\ll n$) of the crystal, i.e. $m\geq Cs$ for some constant $C>0$. The value of $C>0$ will be numerically estimated in Section \ref{sec:results}.

\section{Crystalline Structure Reconstruction Methodology}
\label{sec:method}
In order to solve \eqref{eq:optiproblem}, this work adapted the Sparse Phase Retrieval via Smoothing Function (SPRSF) method introduced in \cite{8410803}. This algorithm claims to have better performance in terms of the number of measurements compared with recent approaches in the state-of-the-art. SPRSF uses a special mapping called smoothing function, which is useful to eliminate the non-smoothness of $h(\cdot)$. Specifically, the optimization problem solved by SPRSF is formulated as
\begin{align}
\hspace{-8em}\min_{\mathbf{x}\in \mathbb{C}^{n}} \hspace{1em} f(\mathbf{x}) &= \frac{1}{m}\sum_{i=1}^{n}\sum_{r=1}^{R}\sum_{p=1}^{P}\left(\sqrt{g^{i}_{p,r}} - \varphi_{\mu}(\lvert \langle\mathbf{b}^{r}_{p,i},\mathbf{x}\rangle \rvert)\right)^{2},
\nonumber\\
\text{ s.t }\hspace{1em}\lVert\mathbf{Fx} \rVert_{0}&\leq s
\label{eq:basicProblem}
\end{align}
where $\varphi_{\mu}(w)$ is defined as $\varphi_{\mu}(w) = \sqrt{w^2+\mu^2}$. The proposed reconstruction algorithm consists on two stages as follows:\vspace{-0.5em}
\begin{itemize}
	\item \textit{Initialization step: } this procedure consists on estimating the crystalline structure $\mathbf{x}$ as the leading eigenvector of a carefully designed matrix.\vspace{-0.5em}
	\item \textit{Refining step: } the outcome of the first step is refined upon a sequence of Wirtinger gradient iterations.\vspace{-0.5em}
\end{itemize}
These two stages are summarized in Algorithm \ref{alg:smothing}.\vspace{-0.5em}
\begin{algorithm}[H]
	\small
	\caption{Crystalline reconstruction algorithm}
	\label{alg:smothing}
	\begin{algorithmic}[1]
		\State {\textbf{Input: } Data $\{(\mathbf{b}^{r}_{p,i};g^{i}_{p,r})\}$. The step size $\tau\in (0,1)$, control variables $\gamma,\gamma_{1}\in (0,1)$, $\mu^{(0)}\in \mathbb{R}_{++}$, number of iterations $T$  and the sparsity $s$.}
		\Statex{}
		\State{\textbf{Initialization: } $\hat{\mathcal{J}}$ set of $s$ largest indices of $\displaystyle\left\{\frac{1}{m}\sum_{i=1}^{n}\sum_{p=1}^{P}\sum_{r=1}^{R}g^{i}_{p,r} \lvert (\mathbf{b}^{r}_{p,i})_{q} \rvert^{2}\right\}_{1\leq q\leq n}$. Let $\tilde{\mathbf{z}}^{(0)}$ be the leading eigenvector of the matrix $\displaystyle\mathbf{H} := \frac{1}{m} \sum_{i=1}^{n}\sum_{p=1}^{P}\sum_{r=1}^{R} g^{i}_{p,r}(\mathbf{b}^{r}_{p,i})_{\hat{\mathcal{J}}}(\mathbf{b}^{r}_{p,i})_{\hat{\mathcal{J}}}^{H}\mathbf{1}_{\{g^{i}_{p,r} \leq \alpha_{y}^{2}\phi^{2}\}}$, where $\alpha_{y}=3$ and $\displaystyle\phi^{2}=\frac{1}{m}\sum_{i=1}^{n}\sum_{p=1}^{P}\sum_{r=1}^{R}g^{i}_{p,r}$}.
		\Statex{}
		\State{$\displaystyle\mathbf{z}^{(0)}\leftarrow \mathbf{F}^{H}\left(\sqrt{\frac{n}{m}\phi^{2}}\right)\tilde{\mathbf{z}}^{(0)}$}
		\Statex{}
		\For{$t=0:T-1$} 
		\State{$\tilde{\mathbf{z}}^{(t+1)} = \mathcal{H}_{s}\left(\mathbf{F}\left(\mathbf{z}^{(t)} - \tau\partial f(\mathbf{z}^{(t)},\mu^{(t)})\right)\right)$}
		\State{$\mathbf{z}^{t+1}\leftarrow \mathbf{F}^{H} \tilde{\mathbf{z}}^{(t+1)}$}
		\Statex{}
		\If {$\lVert \partial f\left(\mathbf{z}^{(t+1)},\mu^{(t)} \right) \rVert_{2} \geq \gamma \mu^{(t)} $}
		\State $\mu^{(t+1)}=\mu^{(t)}$
		\Else
		\State $\mu^{(t+1)}= \gamma_1 \mu^{(t)}$
		\EndIf
		\EndFor
		\State{\textbf{end}}
		\State{\textbf{Output: } $\mathbf{z}^{(T)}$}
	\end{algorithmic}
\end{algorithm}	\vspace{-1em}

Algorithm \ref{alg:smothing} requires the sampling vectors and the acquired coded diffraction patterns as modeled in \eqref{eq:modelfinal} (Line 1). The initialization step is presented in Lines 2-3. Further, a thresholding step is calculated in Line 5, where the operators $\mathcal{H}_{s}(\mathbf{w})$ set all the entries in the vector $\mathbf{w}\in \mathbb{C}^{n}$ to zero, except its $s$ largest absolute values. Additionally if the condition in Line 7 is not satisfied the smoothing parameter $\mu$ is updated in Line 9 to obtain a new point. Remark that each vector $\partial f(\mathbf{z}^{(t)},\mu^{(t)})$ in Algorithm \ref{alg:smothing} is calculated using the Wirtinger derivative \cite{hunger2007introduction} as
\begin{align}
		\partial f\left(\mathbf{z}^{(t)}, \mu^{(t)}\right) = \frac{2}{m}&\sum_{i=1}^{n}\sum_{r=1}^{R}\sum_{p=1}^{P} \left( (\mathbf{b}^{r}_{p,i} )^{H}\mathbf{z}^{(t)}- \right.\nonumber \\
		& \left.\sqrt{g^{i}_{p,r}}\frac{(\mathbf{b}^{r}_{p,i} )^{H}\mathbf{z}^{(t)}}{\varphi_{\mu^{(t)}}(\lvert \langle\mathbf{b}^{r}_{p,i},\mathbf{z}^{(t)}\rangle \rvert)} \right)\mathbf{b}^{r}_{p,i}.
	\label{eq:wirtinger_derivative}
\end{align}
The theoretical guarantees of convergence of this algorithm are demonstrated in \cite{8410803}.

\section{Simulations and Results}
\label{sec:results}
The performance of the proposed algorithm is presented. Three different tests are performed: first, the empirical success of the proposed method is analyzed, among 100 trial runs. The second examines the reconstruction performance of Algorithm \ref{alg:smothing} for recovering the crystalline structure. The third test studies the stability behavior of the recovery algorithm under additive noise.

The default values of the parameters of Algorithm \ref{alg:smothing} were determined using a cross-validation strategy. They were fixed as $\tau=0.3$, $\gamma=0.8$, $\gamma_1=0.5$, $\mu^{(0)}=60$ and $T=800$. The performance metric used is $$\text{relative error :=}\frac{dist(\mathbf{z},\mathbf{x})}{\lVert \mathbf{x} \rVert_{2}},$$ where $dist(\mathbf{z},\mathbf{x})$ is defined as $$\textit{dist}(\mathbf{z},\mathbf{x})= \underset{\theta\in[0,2\pi)}{min}\lVert\mathbf{x} e^{-j\theta}-\mathbf{z}\rVert_{2},$$ with $j=\sqrt{-1}$. The simulated coded aperture was a block-unblock ensemble as used in our previous works \cite{pinilla2018coded,pinilla2018coded1}. All simulations were implemented in Matlab R2019a on an Intel Core i7 3.41Ghz CPU with 32 GB RAM. 

\begin{figure}[h]
	\centering\includegraphics[width=0.8\linewidth]{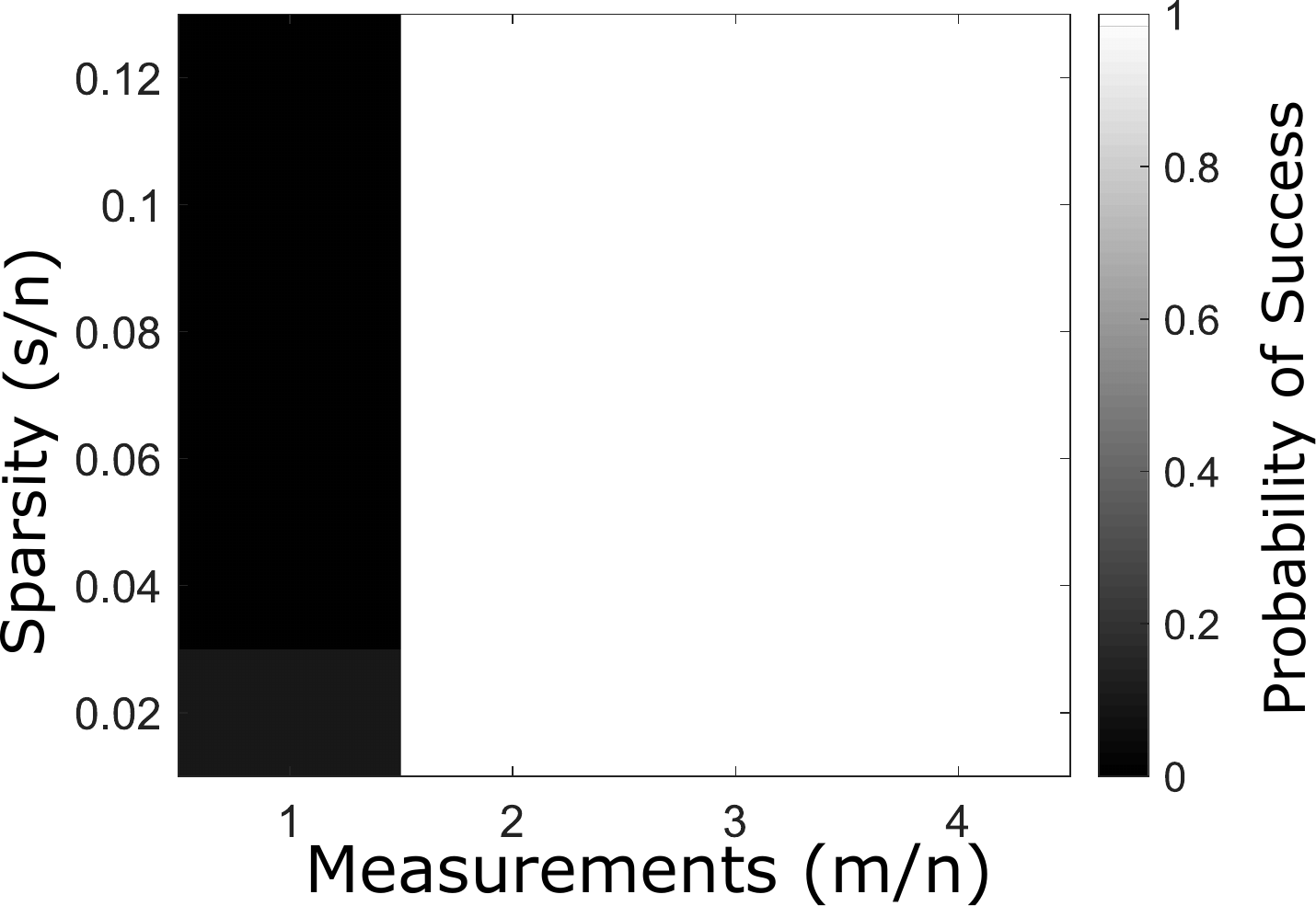}
	\caption{Empirical success rate of Algorithm \ref{alg:smothing} when the sparsity $s$ and the number of measurements $m$ are varied. Each value is determined over 100 trial runs when a relative error of $10^{-5}$ is reached. The white and black colors represent 100\% and 0\% probability of success, respectively. }
	\label{fig:prob}
\end{figure}

To examine the recovery success rate against the level of sparsity $s$ and the number of measurements $m$ under a noiseless scenario, we randomly simulate 1000 crystalline structures. The success rate is determined over 100 trial runs for each crystalline structure when a relative error of $10^{-5}$ is reached. The results are shown in Fig. \ref{fig:prob}, where the success rate is plotted in gray scale when white and black represent 100\% and 0\% probability of success, respectively. 
\begin{figure*}
	\centering\includegraphics[width = 1\linewidth]{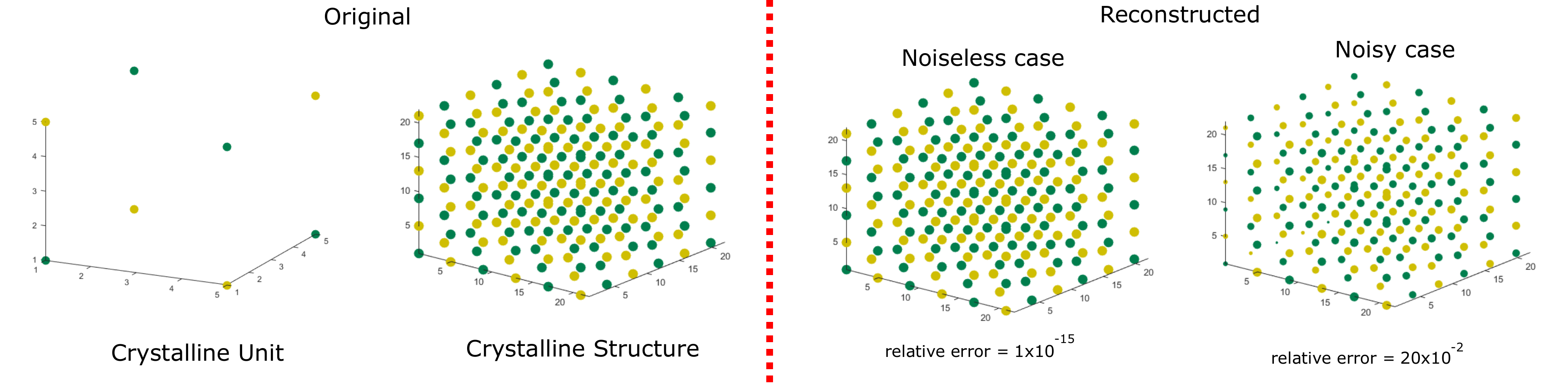}
	\caption{Returned crystalline structure using Algorithm \ref{alg:smothing} for both noiseless and noisy scenarios when $m/n=2$. For the noisy case the SNR = 30dB. The simulated crystal is NaCl.}
	\label{fig:crystal}
\end{figure*}

From Fig. \ref{fig:prob}, it can be concluded that the proposed algorithm is able to estimate the crystalline structure when $m/n\geq 2$ for all the tested sparsity levels. This result implies that the proposed acquisition system, along with its reconstruction approach, is able to estimate the crystal structure using $50\%$ less measurements than the state-of-the-art methods that require $m/n\geq 4$ \cite{conca2015algebraic,bendory2017fourier}. Thus, considering the fact that $s\ll n$, then the constant $C$ of Theorem \ref{theo:guarantee} is bounded from above as $C\leq 2$. Observe that in practice we can achieve this reduction by choosing the number of sensing distances as $P=2$.

Finally, the reconstruction accuracy of Algorithm \ref{alg:smothing} for both noiseless and noisy scenarios is analyzed. The tested crystalline structure is the sodium chloride (NaCl) as illustrated in Fig. \ref{fig:crystal}. Specifically the green spheres in Fig. \ref{fig:crystal} models the Cl atoms and the yellow ones the sodium atoms. For the noisy case, we study the stability behavior of Algorithm \ref{alg:smothing} under additive white noise with a Signal-to-Noise-Ratio (SNR) fixed as $SNR = 30dB$ where SNR$ = 20\log(\lVert \mathbf{g} \rVert_{2}/(m\sigma))$ with $\sigma$ the variance of the noise. The number of measurements $m$ used for this experiments satisfies $m/n=2$. The attained reconstructions are shown in Fig. \ref{fig:crystal} suggesting the effectiveness of the proposed method to estimate the crystalline structure from both noiseless and noisy measurements.\vspace{-0.5em}

\section{Conclusion and Discussion}
This work presented an algorithm to recover the 3D structure of a crystal, under a system that records coded diffraction patterns. Our approach takes advantage of the fact that the crystalline structure can be sparsely represented in the Fourier domain to significantly reduce the number of measurements. Simulations show that our approach is able to reconstruct the crystalline structure with up to $50\%$ less amount of measurements compared with the traditional methods in the state-of-the-art. Also, the results suggest that the proposed approach allows reconstructing the crystalline structure even in noisy scenarios.

To implement the proposed acquisition system in Fig. \ref{fig:proposed} a block-unblock coded aperture is feasible \cite{pinilla2018coded,pinilla2018coded1}. Specifically, the blocking elements of these coded apertures can be fabricated using tungsten, since this material can stop an x-ray beam, resulting in low fabrication costs \cite{maccabe2013snapshot,brady2013coded,WinNT}.

\appendix
\section{Proof of Theorem \ref{theo:guarantee}}
\label{app:proofTheogua}
To prove Theorem \ref{theo:guarantee} we divide it into two parts. First, we prove the right inequality in \ref{rip}, and then as a second part we prove the left inequality. Thus, let $\mathbf{W}\in \mathcal{T}_{\mathbf{x}}$. As $\mathbf{W}$ has rank at most two, we can choose normalized vectors $\mathbf{u},\mathbf{v}\in \mathbb{C}^{n}$ such that $\mathbf{W} = \lambda_{1}\mathbf{u}\mathbf{u}^{H}+\lambda_{2}\mathbf{v}\mathbf{v}^{H}$. Then considering the definition of the linear map $\mathcal{B}$ in \eqref{eq:map} we have that
\begin{align}
	\lVert \mathcal{B}(\mathbf{W}) \rVert_{1} &= \sum_{i=1}^{n}\sum_{r=1}^{R}\sum_{p=1}^{P}\left\lvert \lambda_{1}\lvert \langle\mathbf{b}^{r}_{p,i},\mathbf{u} \rangle \rvert^{2} + \lambda_{2}\lvert \langle\mathbf{b}^{r}_{p,i},\mathbf{v} \rangle \rvert^{2} \right\rvert \nonumber\\
	&\leq \sum_{i=1}^{n}\sum_{r=1}^{R}\sum_{p=1}^{P}\lvert \lambda_{1}\rvert\lvert \langle\mathbf{b}^{r}_{p,i},\mathbf{u} \rangle \rvert^{2} + \lvert\lambda_{2}\rvert\lvert \langle\mathbf{b}^{r}_{p,i},\mathbf{v} \rangle \rvert^{2} \nonumber\\
	&=\lvert \lambda_{1}\rvert\lVert \mathbf{B}\mathbf{u} \rVert_{2}^{2} + \lvert \lambda_{2}\rvert\lVert \mathbf{B}\mathbf{v} \rVert_{2}^{2}\leq \lVert \mathbf{W}\rVert_{1}\lVert \mathbf{B} \rVert_{\infty}^{2},
	\label{eq:inequality1}
\end{align}
in which the first and second inequalities are obtained using the triangular inequality, and matrix $\mathbf{B}$ as defined in Theorem \ref{theo:guarantee}. Further, considering definition of matrix $\mathbf{B}$ we have that
\begin{align}
	\mathbf{B}^{H}\mathbf{B} & = \sum_{i=1}^{n}\sum_{r=1}^{R}\sum_{p=1}^{P} \mathbf{F}^{H}\mathbf{S}_{r}\overline{\hat{\mathbf{D}}}\mathbf{F}\overline{\mathbf{T}}(z_{p})\mathbf{f}_{i}\mathbf{f}_{i}^{H}\mathbf{T}(z_{p})\mathbf{F}^{H}\hat{\mathbf{D}}\mathbf{S}_{r}\mathbf{F} \nonumber\\
	&=\overline{\hat{\mathbf{D}}}\hat{\mathbf{D}},
	\label{eq:isotropic1}
\end{align}
since $\mathbf{F}^{H}\mathbf{F} = \mathbf{F}\mathbf{F}^{H}=\mathbf{I}$, $\mathbf{T}(z_{p})$ is a diagonal orthogonal matrix, and $\sum_{r=1}^{R}\mathbf{S}_{r}\mathbf{S}_{r}=\mathbf{I}$. Thus, using the fact that the admissible random variable $d$ is assumed $|d|\leq 1$ we have from \eqref{eq:isotropic1} that $\mathbf{B}$ is an isotropic subgaussian matrix \cite{vershynin2010introduction}. Then, from Theorem 5.39 in \cite{vershynin2010introduction} we have that
\begin{align}
	\mathcal{P}\left(\left\lVert \mathbf{B} \right\rVert_{\infty} \geq \sqrt{m}+C\sqrt{s}+t\right)\leq 2e^{-c_{0}t^{2}},
	\label{eq:concentration}
\end{align}
for constants $c_{0},C>0$ and any $t>0$. Then, taking $m\geq C^{2}\epsilon^{-2}s$ and $t=\sqrt{m}\epsilon$ for any $\epsilon\in (0,1/2)$, we have from \eqref{eq:concentration} that 
\begin{align}
	\mathcal{P}\left(\frac{1}{m}\lVert \mathbf{B} \rVert^{2}_{\infty} \leq 1+\delta\right)\leq 1-2e^{-c_{0}m\epsilon^{2}},
\end{align}
for $\delta = 2\epsilon$. Thus, we conclude that
\begin{align}
	\frac{1}{m}\lVert \mathcal{B}(\mathbf{W}) \rVert_{1} \leq (1+\delta)\lVert \mathbf{W}\rVert_{1},
	\label{eq:rightside}
\end{align}
for any $\delta\in (0,1)$.

On the other hand, from \eqref{eq:inequality1} we can also conclude that
\begin{align}
	\lVert \mathcal{B}(\mathbf{W}) \rVert_{1} &\geq \sum_{i=1}^{n}\sum_{r=1}^{R}\sum_{p=1}^{P} \lambda_{1}\lvert \langle\mathbf{b}^{r}_{p,i},\mathbf{u} \rangle \rvert^{2} + \lambda_{2}\lvert \langle\mathbf{b}^{r}_{p,i},\mathbf{v} \rangle \rvert^{2}\nonumber\\
	&=  \lambda_{1}\lVert \mathbf{B}\mathbf{u} \rVert_{2}^{2} +  \lambda_{2}\lVert \mathbf{B}\mathbf{v} \rVert_{2}^{2} \nonumber\\
	&= \left(\lambda_{1} + \lambda_{2}\right) = \lVert \mathbf{W}\rVert_{1},
	\label{eq:leftside}
\end{align}
in which the third equality comes from observation in \eqref{eq:isotropic1}, using that $\mathbf{W}$ is assumed to be positive semidefinite. Thus, we have that
\begin{align}
	\frac{1}{m}\lVert \mathcal{B}(\mathbf{W}) \rVert_{1} \geq \frac{1}{m}\left(1-\delta\right)\lVert \mathbf{W}\rVert_{1},
	\label{eq:leftside1}
\end{align}
for any $\delta \in (0,1)$. Thus, combining \eqref{eq:rightside} and \eqref{eq:leftside1} the result holds.

\bibliography{apssamp}

\end{document}